\documentclass[10pt]{article}
\usepackage{conf_iap}
\usepackage{psfig}
\newcommand{\boom}{{\sc Boomerang}}
\newcommand {\acbar}{{\sc Acbar}}
\newcommand {\cobedmr}{{\sc COBE-DMR}}
\newcommand{\apjl}{ApJL\,\,}
\newcommand{\prd}{PRD\,\,}
\begin{document}
\heading{%
%
Limits on the Dark Energy Parameters from Cosmic Microwave Background
 experiments  \\
%
} 
\par\medskip\noindent
\author{%
Dmitri Pogosyan$^{1}$, J.~Richard~Bond$^{2}$, Carlo~Contaldi$^{2}$
}
\address{%
Physics Department, University of Alberta, Edmonton, AB, T6G 2J1, Canada
}
\address{%
CITA, University of Toronto, Toronto, ON, M5S 3H8, Canada
}

\begin{abstract}
Full suite of the present day Cosmic  Microwave background (CMB) data, when
combined with weak prior information on the Hubble constant and the
age of the Universe, or the Large-Scale structure, provides strong indication
for a non-zero density of the vacuum-like dark energy in our universe.
This result independently
supports the conclusions from Supernovae Ia (SN1a) data. 
When the model parameter space is extended to allow for the range of the
equation of state parameter $w_Q$ for the dynamical field $Q$
which gives rise to dark energy,
the CMB data is found to give a weak upper bound $w_Q < -0.4$ 
at $95\%$ CL,
however combined with SN1a data it strongly favours $w_Q < -0.8$,
consistent with $\Lambda$-term like behaviour. 
\end{abstract}
\vspace{1cm}

Measurements of anisotropies in the cosmic microwave background (CMB)
radiation now tightly constrain the nature 
and composition of our universe.  High signal-to-noise
detections of primordial anisotropies have been made 
at angular scales ranging from the quadrupole~\cite{bennett96} to
as small as several arcminutes~\cite{mason02,pearson02,dawson02}.
Within the context of models with adiabatic initial 
perturbations, as are generally predicted by inflation,
these measurements have been used in combination with
various other cosmological constraints to estimate the values
of many important cosmological parameters.  Combining their 
CMB data with weak cosmological constraints such as a very loose 
prior on the Hubble constant $H_0$, various teams have made 
robust determinations of several parameters, including the 
total energy density of the universe $\Omega_{tot}$, 
the density of baryons $\omega_b = \Omega_b h^2, h=H_0/100\,
{\mathrm km/\mathrm s/\mathrm Mpc}$,
and $n_s$, the spectral index of primordial perturbations
~\cite{lange01,balbi00,pryke01,netterfield01}.

In Table~1 we present some of our results for
the cosmological parameters together with $68\%$ CL errors,
obtained from the latest suite of CMB observations.
Full details can be found in \cite{ruhl02,gold02}.
Our analysis has included \acbar~\cite{kuo02},Archeops~\cite{benoit02},
\boom~\cite{ruhl02}, CBI~\cite{pearson02}, DASI~\cite{halverson01},
MAXIMA~\cite{lee01}, VSA~\cite{scott02} and \cobedmr~\cite{bennett96} data. 
Combining these highly consistent datasets allows for a very
tight determination of the baryon and dark
matter physical densities $\omega_b$ and $\omega_{cdm}$ and the slope of
scalar perturbations $n_s$, even with no prior 
restrictions on the parameters.

\begin{center}
\begin{tabular}{lllllll}
\multicolumn{7}{l}{{\bf Table 1.} Cosmological parameters from combination of CMB experiments } \\
\\\hline
\multicolumn{1}{c}{Priors} &
\multicolumn{1}{c}{$\Omega_{tot}$} &
\multicolumn{1}{c}{$n_s$} &
\multicolumn{1}{c}{$\omega_b$} &
\multicolumn{1}{c}{$\omega_{cdm}$} &
\multicolumn{1}{c}{$\Omega_{\Lambda}$} &
\multicolumn{1}{c}{$\tau_c$}
\\
\hline
\\
no prior
& $1.2^{0.11}_{0.10}$ 
& $0.96^{0.04}_{0.04}$ 
& $0.022^{0.002}_{0.002}$ 
& $0.11^{0.02}_{0.02}$ 
& $0.24^{0.21}_{0.15}$ 
& $<0.33$ 
\\
\\
wk
& $1.06^{0.04}_{0.04}$ 
& $0.95^{0.04}_{0.03}$ 
& $0.022^{0.002}_{0.002}$ 
& $0.12^{0.02}_{0.02}$ 
& $0.56^{0.13}_{0.14}$ 
& $<0.31$ 
\\
wk+LSS
& $1.05^{0.03}_{0.03}$ 
& $0.96^{0.03}_{0.03}$ 
& $0.022^{0.002}_{0.002}$ 
& $0.11^{0.02}_{0.02}$ 
& $0.66^{0.09}_{0.09}$ 
& $<0.27$ 
\\
wk+SN
& $1.03^{0.03}_{0.03}$ 
& $0.96^{0.03}_{0.03}$ 
& $0.023^{0.002}_{0.002}$ 
& $0.10^{0.02}_{0.01}$ 
& $0.70^{0.04}_{0.05}$ 
& $<0.39$ 
\\
wk+LSS+SN
& $1.03^{0.03}_{0.03}$ 
& $0.97^{0.03}_{0.03}$ 
& $0.023^{0.002}_{0.002}$ 
& $0.10^{0.02}_{0.01}$ 
& $0.70^{0.04}_{0.04}$ 
& $<0.37$ 
\\
\\
$\Omega_{tot}=1$ plus &&&&&&\\
wk
& (1.00) 
& $0.96^{0.03}_{0.03}$ 
& $0.022^{0.002}_{0.002}$ 
& $0.12^{0.02}_{0.02}$ 
& $0.72^{0.07}_{0.08}$ 
& $<0.17$ 
\\
wk+LSS
& (1.00) 
& $0.97^{0.03}_{0.03}$ 
& $0.022^{0.002}_{0.002}$ 
& $0.12^{0.01}_{0.01}$ 
& $0.70^{0.05}_{0.04}$ 
& $<0.17$ 
\\
wk+SN
& (1.00) 
& $0.96^{0.03}_{0.03}$ 
& $0.022^{0.002}_{0.002}$ 
& $0.12^{0.01}_{0.01}$ 
& $0.71^{0.05}_{0.04}$ 
& $<0.18$ 
\\
wk+LSS+SN
& (1.00) 
& $0.97^{0.02}_{0.03}$ 
& $0.022^{0.002}_{0.002}$ 
& $0.12^{0.01}_{0.01}$ 
& $0.70^{0.04}_{0.04}$ 
& $<0.18$ 
\\
\\
\hline
\end{tabular}
\end{center}

Here we shall focus on the parameters
describing the vacuum (or dark) energy, which
may dominate the present day energy balance. In the minimal parameter
set of Table~1 the vacuum energy is assumed to behave as $\Lambda$-term
being described just by its density parameter $\Omega_{\Lambda}$.
Supernovae~Ia results \cite{perlmutter99} provide a strong indication that
$\Omega_{\Lambda} > 0$.

\begin{figure}[ht]
\centerline{
\psfig{figure=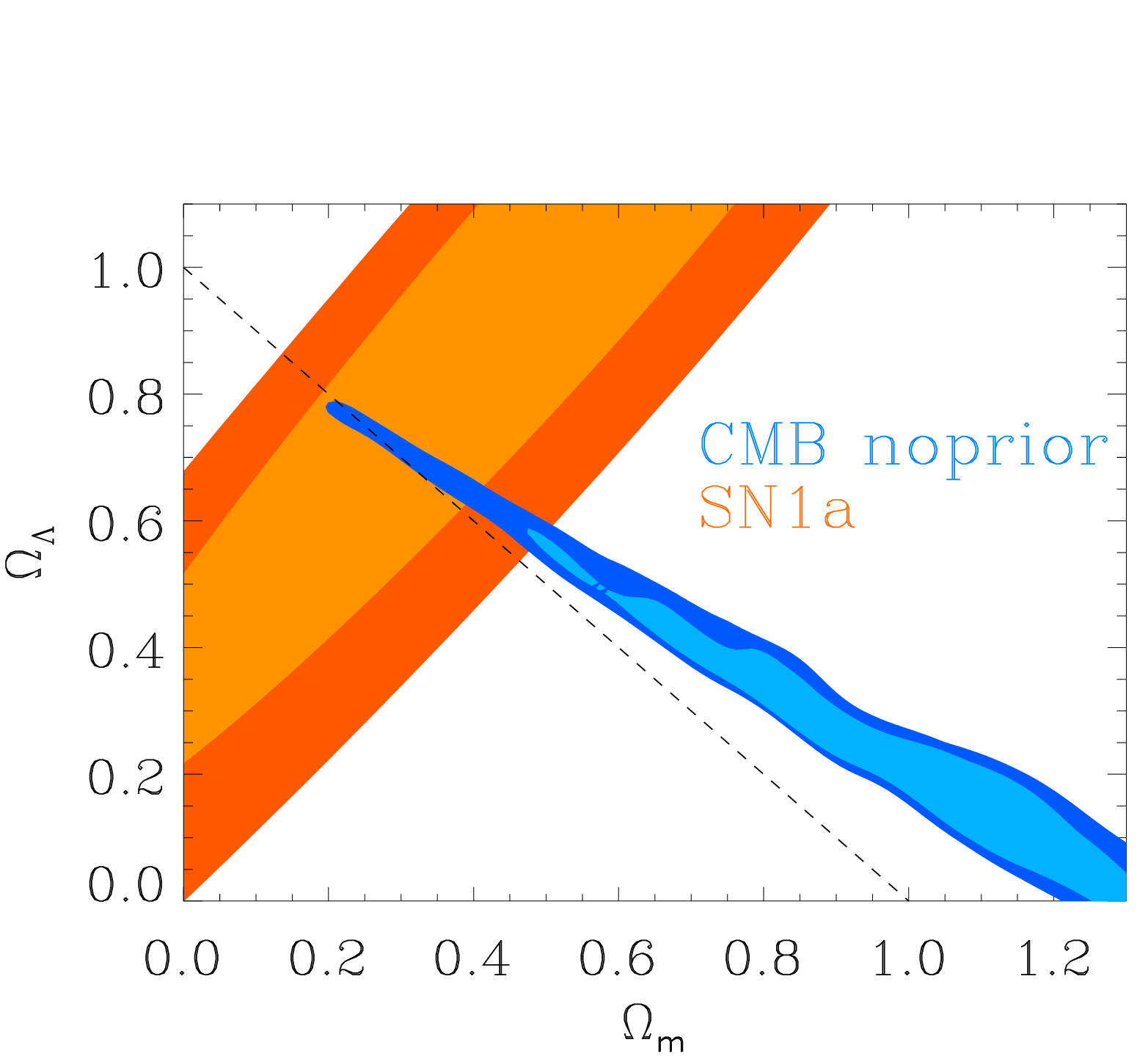,height=4.7cm}
\psfig{figure=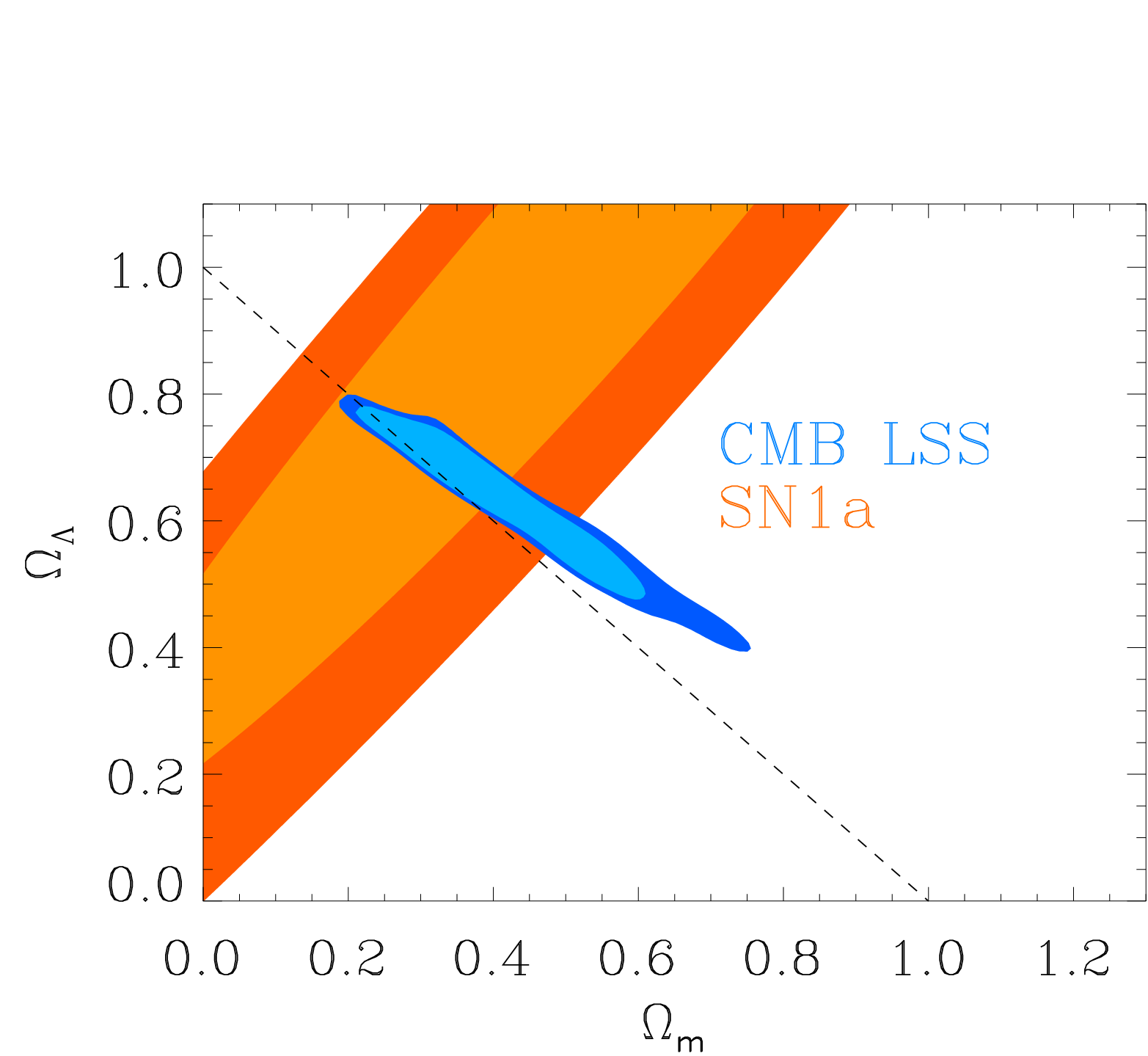,height=4.7cm}
}
\centerline{
\psfig{figure=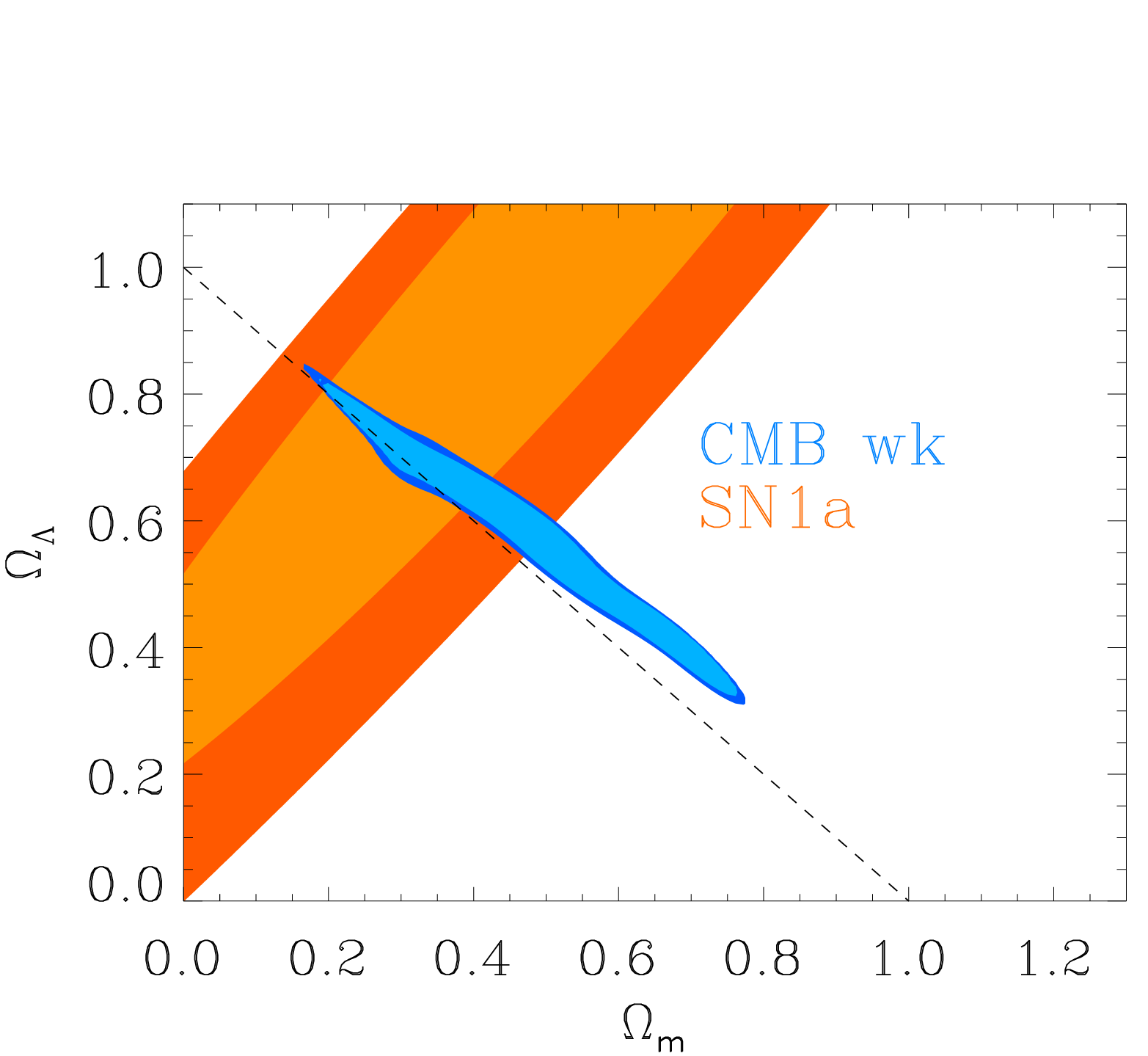,height=4.7cm}
\psfig{figure=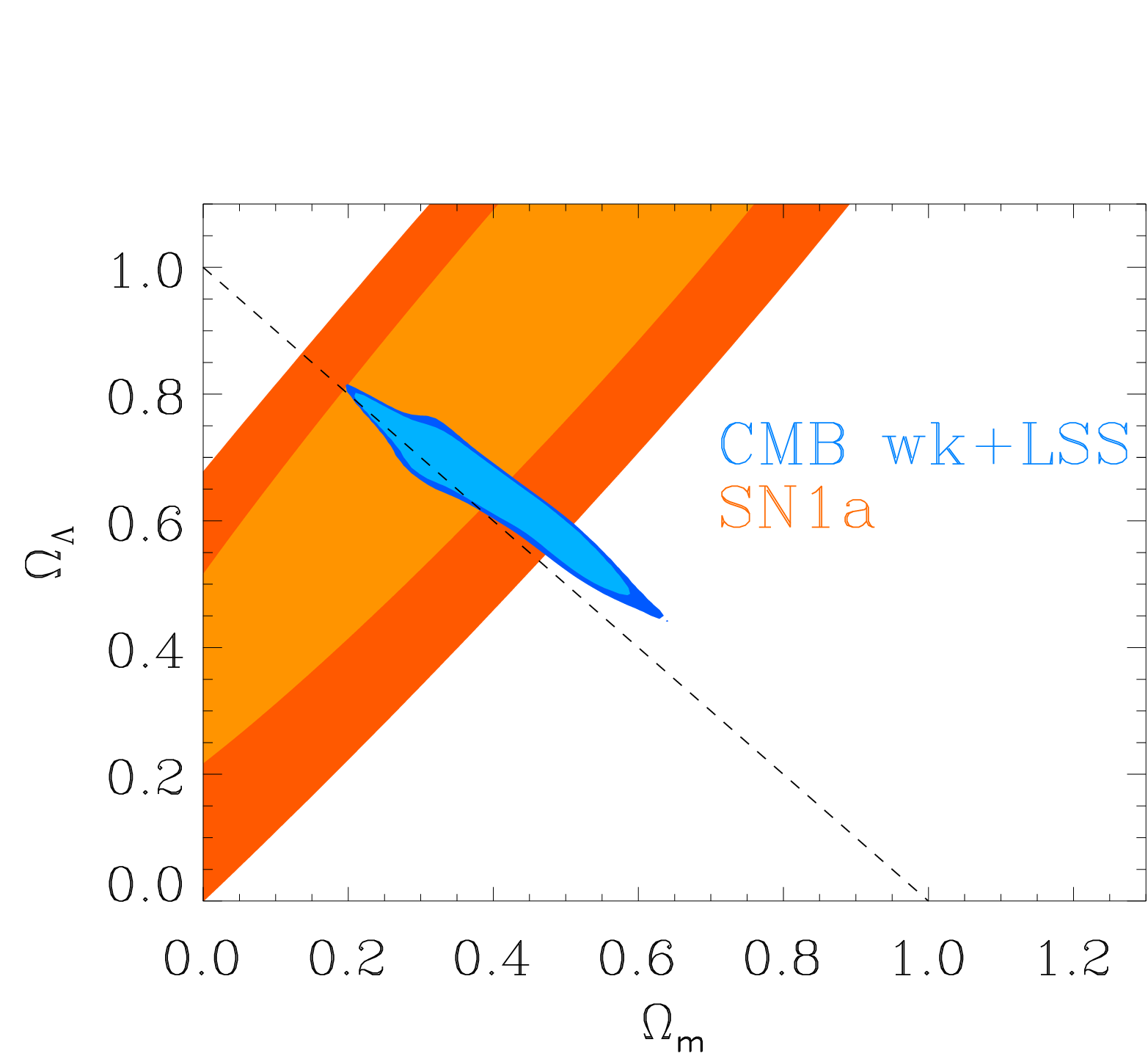,height=4.7cm}
}
\caption[]{Contour regions of the marginalized likelihood 
in $\Omega_m=1-\Omega_{tot}-\Omega_{\Lambda}$ versus $\Omega_{\Lambda}$
plane. Results from CMB data
(blue) are superimposed on the contours from SN1a data (orange).
Shaded areas correspond to $ 68\% $ and $95\% $ confidence regions
in the Gaussian limit. Degeneracy of the likelihood along the
line of the constant angular size of the sound horizon
exhibited by CMB data (upper left) is reduced by either imposing a weak prior
on the Hubble constant and age (bottom left) or by the Large Scale Structure
restrictions (upper right) or both (bottom right).}
\label{Fig:OlOm}
\end{figure}

The non-prior CMB analysis exhibits persistent degeneracy in
the $\Omega_{tot} - \Omega_{\Lambda}$ space (see Figure 
\ref{Fig:OlOm}), which extends along the direction of the constant angular size
of the sound horizon at recombination --- from closed models
$\Omega_{tot} \sim 1.2 $ with little if any vacuum energy contribution,
to flat models
$\Omega_{tot} \approx 1, ~ \Omega_\Lambda \approx 0.7$.
The degeneracy is reduced and the high likelihood region is concentrated near
flat models with significant vacuum energy as soon as 
even weak prior restrictions (``wk'' in Table~1) are
imposed on the Hubble constant 
$0.45 < h < 0.9$ and the age of the universe $t > 10$\,Gyr,
or when CMB data is combined with the Large-Scale Structure (LSS) information, 
namely the amplitude $\sigma_8$ and the slope of density power spectrum
on galaxy cluster scale (see \cite{gold02} for the details of the LSS prior).
Thus, present day CMB data provides new independent indication
for non zero vacuum density.
This conclusion is significantly strengthened by the new data obtained
this year, in particular, by
the Archeops team which has measured CMB temperature fluctuations
in the intermediate angular range $l\sim 20-150$, only weakly
constrained by previous data.

We note that
the data indicates that the optical depth due to late-time reionization
in the Universe is not large, with $95\%$ upper limit being
in the range $\tau_c < 0.3-0.4$. The exact value, however, is not yet robust.

The plausible interpretation of the dark energy is that it is the energy of a
``quintessence'' scalar field $Q$, dynamical in origin, thus,  
$\Omega_{\Lambda}=\Omega_Q$.
A popular phenomenology is to describe additional complexity by
one more parameter $w_Q=P_Q/\rho_Q$, the ratio of the pressure $P_Q$
and the density $\rho_Q$ of the field which serves as an effective
equation of state.  As long as we do not have a
pure $\Lambda$ term with $w_Q=-1$, the $w_Q$ is expected to
vary in time, but the precision of CMB data does not yet warrant such
detailed parametrization.
We consider $w_Q$ constant and restict our
attention to inflation motivated flat $\Omega_{tot}=1$ cosmological models
(this prior does not affect other parameters, cf. lower part of Table~1).
As another simplification we neglect the effect of late-time
perturbations in the $Q$ field, which may affect CMB signal at the
largest scales.
This effect is dependent in detail on the exact model for quintessence potential
(one should be cautioned not to extend the equation
of state analogy too far, and treat perturbations in the $Q$ component
hydrodynamically).

\begin{figure}[t]
\centerline{
\psfig{figure=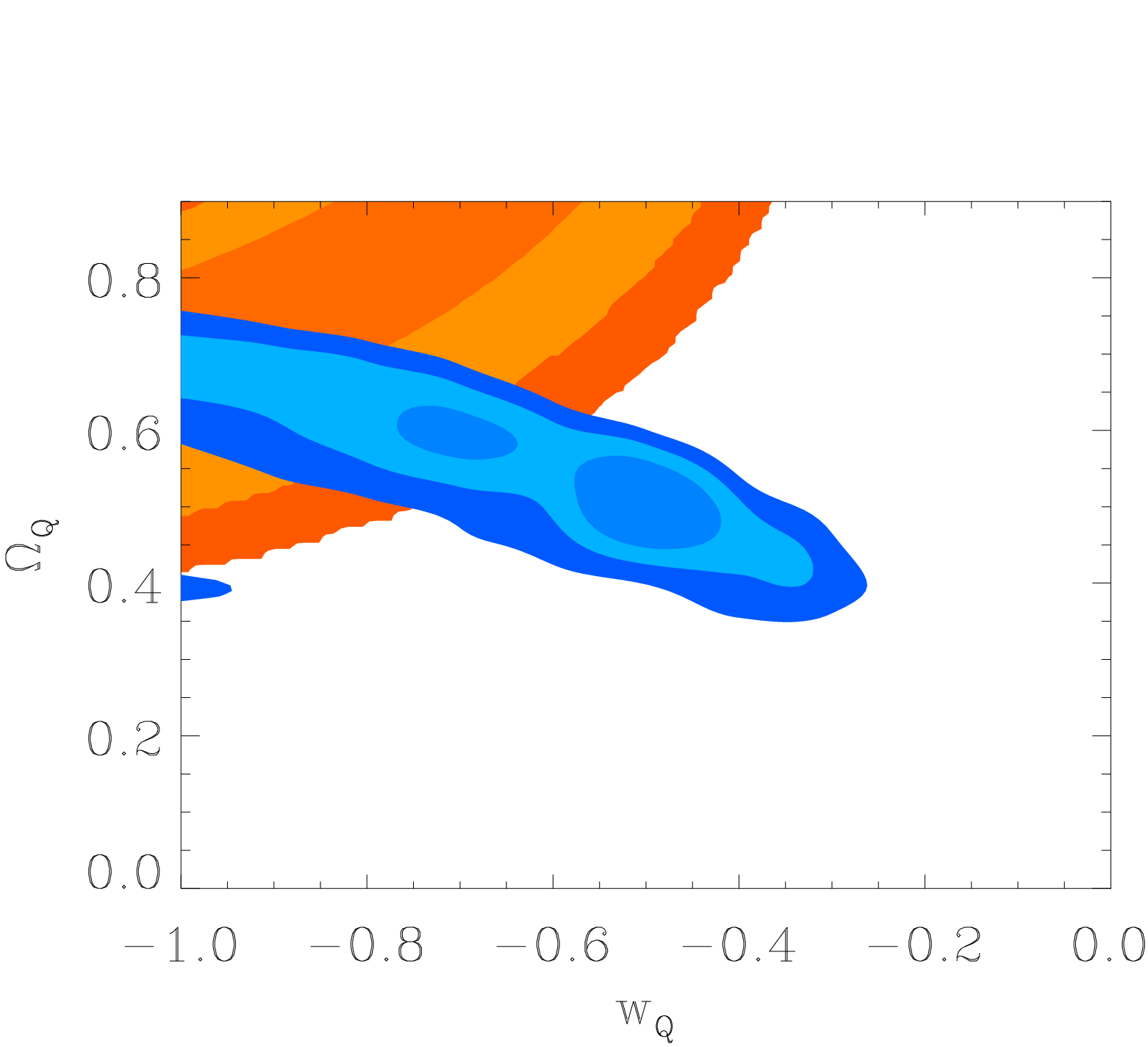,height=4.7cm}
\psfig{figure=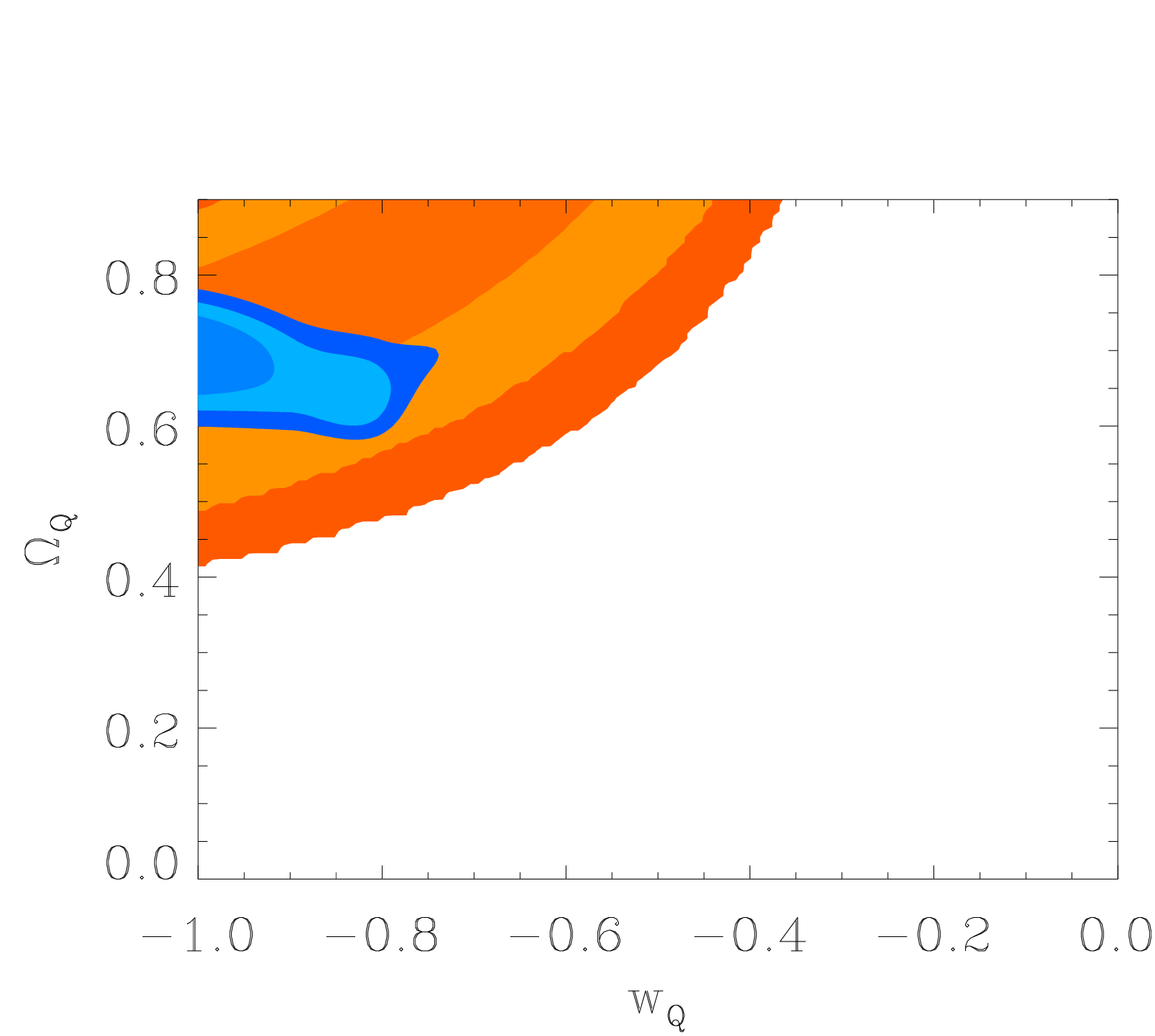,height=4.7cm}
}
\caption[]{Contours of marginalized likelihood for the
parameters of the $Q$-field $\Omega_Q, w_Q$ in the class of spatially
flat $\Omega_{tot}=1$ cosmological models. Left -- CMB with weak prior,
right -- joint CMB and SN1a
results (blue) are superimposed on SN1a limits (orange).
Shaded regions correspond
to $68\%, 95\%$ (the lightest colour) and $99\%$ CL
limits in the Gaussian approximation.
}
\label{Fig:Oqwq}
\end{figure}

Figure~\ref{Fig:Oqwq} shows that CMB alone does not yield
useful restriction on the $w_Q$ ($w_Q < 0.4$ at 95\% CL)
but its combination with SN1a data prefers at $95\%$ CL 
the low values $w_Q < -0.8$, an improvement to the limit reported in \cite{bond02}.

\acknowledgements{We are grateful to \boom, CBI and \acbar
~collaborations with whom 
much of the work reviewed here has been done.
The computational facilities at Toronto are funded by the
Canadian Fund for Innovation.
}

\vfill
\end{document}